\documentclass[aps,twocolumn,showpacs]{revtex4}
\usepackage{amsmath}
\usepackage{epsfig}
\usepackage{amssymb}
\usepackage{graphicx}
\usepackage{float}
\usepackage{xcolor}
\begin{document}

\title{Three Dimensional Gravity and Schramm-Loewner Evolution  }

\author{Jing Zhou}
\email{171001005@njnu.edu.cn}
\affiliation{Department of Physics, Hunan City University, Yiyang, Hunan 413000, China}
\affiliation{All-solid-state Energy Storage Materials and Devices Key Laboratory of Hunan Province, Hunan City University, Yiyang 413000, China}

\begin{abstract}
The partition function of three dimensional gravity in the
quantum regime is dual to the Ising model when the central charge $c=1/2$. Mathematically, we show that the three dimensional gravity can be described by Schramm-Loewner Evolution(SLE) with certain $\kappa$. In fact, SLE depends on the parameter $\kappa$ which controls the diffusion of the Brownian motion.
Each value of $c < 1$ corresponds to two values of $\kappa$, which may hint that the three dimensional gravity has two different phases at certain central charge c.
Moreover, phase transition is also discussed in AdS and Ising model.

\end{abstract}


\maketitle

\section{\label{sec:introduction}Introduction}
It is an interesting problem to understand whether three-dimensional gravity
exists as a quantum theory, and to solve it, if it exists. The
action can be written as three-dimensional gravity with a cosmological constant was studied in~\cite{Deser:1983nh,Yin:2007gv}.
In early studies of the theory, many can be special to the case of negative cosmological constant.
It was firstly discovered by Banados, Teitelboim, and Zanelli~\cite{Banados:1992wn,Banados:1992gq}that there exist black hole solutions in three dimensional gravity with negative cosmological constant.
Subsequent work has made it clear that the three-dimensional black holes should be
taken seriously in the AdS/CFT correspondence~\cite{Maldacena:1997re,Witten:1998qj}.
The existence of the BTZ black hole can be considered as a chance for a solvable quantum model.
Unfortunately, the existence nonlinear interactions in four dimensional gravity means that there is no an exact solution of the system. However, if three-dimensional gravity is quantum gravity, there will be many black holes here,

As we know, when $\Lambda > 0$, there is no nonperturbative solution in three-dimensional gravity. The main reason is that the physical system with positive $\Lambda$ is always metastable. Therefore, the pure gravity with $\Lambda > 0$ is meaningless as an exact solvable theory. Now, Let us turn to gravity with $\Lambda = 0$. One can find that there is no S matrix and black hole solution in this theory. Through the simple analysis of the above two cases, we finally study gravity $\Lambda <0$. The three-dimensional gravity with $\Lambda <0$ coupled to additional fields was firstly studied by Brown and Henneaux \cite{Brown:1986nw}. The main result is that the physical Hilbert space obtained in this theory has an action of left and right moving Virasoro algebras with $c_{L} = c_{R} = 3\ell/2G$ \cite{Brown:1986nw}. In fact, one can find that this is the boundary conformal field theory.

The Schramm-Loewner evolution (SLE)~\cite{Schramm:1999rp,Duplantier:1999cz,Schramm:1999cz,Bauer:2006gr,Rohde:2001rm,Kenyon:2015bhs,Gwynne:2015koj,Doyon:2005wy,Dubedat:2015uya,Santachiara:2007oik} is closely related to two-dimensional conformal field theory. Mathematically, SLE describes fractal curves found in critical percolation percolation\cite{Santachiara:2007oik}. In Conformal field theories (CFT) the emphasis is on the algebraic
properties of the critical models dictated by their conformal invariance \cite{Santachiara:2007oik,Najafi}. Therefore, we can find that there are some links between CFT and SLE. Moreover, according to the AdS/CFT correspondence, we can study CFT from the perspective of three-dimensional gravity. However, we find that it is still an open question to understand three-dimensional gravity directly from the aspect of SLE. Motivated by this, it is natural to study the connection between three dimensional gravity and SLE in this work. This may provide a new perspective for us to understand three-dimensional gravity.  The rest of this paper is organized as follows. In Sec.~\ref{sec:02}, we discuss the three dimensional gravity and Ising model. In Sec.~\ref{sec:03}, we mainly focus on phase transition in Ising model and three dimensional gravity. Further discussion and conclusion can be found in Sec.~\ref{sec:04}

\section{Three dimensional gravity and Ising model }\label{sec:02}
Let us now consider gravity where the AdS radius $\ell$ is at Planck scale. Firstly, we focus on the path integral in three dimensional quantum gravity with negative cosmological constant~\cite{Castro:2011zq,Santachiara:2007oik,Dijkgraaf:2000fq,Maldacena:1998bw,Kraus:2006wn}. The path integral of quantum gravity is given by
\begin{equation}
Z_{\text {grav }}=\int_{\partial \mathcal{M}} \mathcal{D} g e^{-c S_{E}[g]}
\end{equation}
Note that the Euclidean action $S_{E}$ of the gravity is closely related to the central charge. When $c < 1$, the partition function of $AdS_{3}$ quantum gravity read~\cite{Castro:2011zq}
\begin{equation}
Z_{\text {grav }}(\tau, \bar{\tau})=\sum_{\gamma \in \Gamma_{e} \backslash S L(2, Z)} Z_{\text {vac }}(\gamma \tau, \gamma \bar{\tau})
\end{equation}
Here $\Gamma_{e}$ is the subgroup.
 $Z_{vac}$ is the vacuum character partition function of  minimal model CFT ~\cite{Castro:2011zq}, namely
\begin{equation}
Z_{\mathrm{vac}}(\tau, \bar{\tau})=\left|\chi_{1,1}(\tau)\right|^{2}
\end{equation}
With ~\cite{Castro:2011zq}
\begin{widetext}
\begin{eqnarray}
\chi_{1,1}= &q^{(1-c) / 24} \frac{(1-q)}{\eta(\tau)} &\left(1+\sum_{k=1}^{\infty}(-1)^{k}\left(q^{h_{1+k(p+1),(-1)^{k}+\left(1-(-1)^{k}\right){p / 2}}} +q^{h_{1, kp+(-1)^{k}+\left(1-(-1)^{k}\right){p / 2}}} \right)\right).
\end{eqnarray}
\end{widetext}
where~\cite{Castro:2011zq}
\begin{equation}
h_{r, s}=\frac{(p r-(p+1) s)^{2}-1}{4 p(p+1)}.
\end{equation}
Then the partition function of this Ising model( $c=\frac{1}{2}$) can be written as~\cite{Castro:2011zq}
 \begin{equation}
Z_{\text {Ising }}(\tau, \bar{\tau})=\left|\chi_{1,1}(\tau)\right|^{2}+\left|\chi_{2,1}(\tau)\right|^{2}+\left|\chi_{1,2}(\tau)\right|^{2}
\end{equation}
Thus one can find~\cite{Castro:2011zq}
\begin{equation}
Z_{grav} = 8Z_{Ising}
\end{equation}
Thus, we find that the partition function of pure quantum gravity at $c=\frac{1}{2}$ is equivalent to that of the Ising model~\cite{Castro:2011zq}.

However, this is not the whole story. It is important to note that the Ising model or CFT is closely related to SLE when the central charge is considered. The random collection of conformal maps $g_{t}(z)$ can be obtained by solving the Loewner equation\cite{Smirnov:2010cz,Santachiara:2007oik}
\begin{equation}
\partial_{t} g_{t}(z)=\frac{2}{g_{t}(z)-\sqrt{\kappa} B_{t}}, \quad g_{0}(z)=z
\end{equation}
Generally, the central charge $c$ plays an important role in determining the scaling limits in conformal field theory (CFT)\cite{Castro:2011zq}.  We can now observe that the SLE parameter and the central charge $c$ are related through the following relationship\cite{Castro:2011zq,Santachiara:2007oik}
\begin{equation}
c=\frac{(6-\kappa)(3 \kappa-8)}{2 \kappa}
\end{equation}
The parameter $\kappa$ describing the boundary SLE can be determined by coupling the SLE process to conformal field theory\cite{Schramm:2004cz}. When the partition function of pure quantum gravity is at $c=\frac{1}{2}$,
we obtain
\begin{equation}
\kappa =3,\frac{16}{3}
\end{equation}
Or in the other word, the three dimensional pure quantum gravity at $c=\frac{1}{2}$ can be described by SLE with $\kappa =3,\frac{16}{3}$. And one should note that the $\kappa$ is the weight of Brown motion.

Following Ref.~\cite{Castro:2011zq}, we calculated $\kappa$ for pure gravity, $E_{6}$ gravity, $SL_{3}$ gravity and $SL_{N}$ gravity. And all the results can be found in table I. It is not difficult to see that for a given gravity, $\kappa$ has two values. This means that gravity has two different phases.

Actually, when
$0 <\kappa \le 4$: $\gamma(t)$ is a simple curve, but it doesn't touch itself.
When $4 < \kappa < 8$: $\gamma(t)$ is not a simple curve. And the curve touches itself.
When $\kappa\ge8$: $\gamma(t)$ is a space filling curve~\cite{Cardy:2005kh}. It has double points, but does not cross itself. The three different types of behaviour are sketched in Fig. 1.
\begin{figure}[H]
\centering
\epsfxsize=3.3in \epsfbox{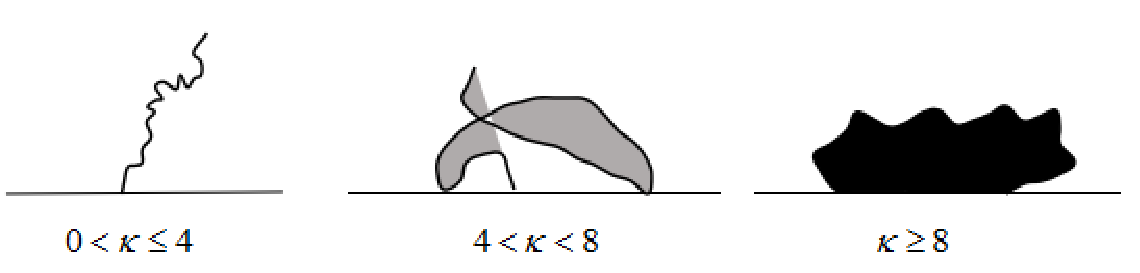} \vspace{-0.1in}
\caption{Three different phases for $\kappa$ whose corresponding SLE partly
show a different behaviour, namely $0 <\kappa \le 4$, $4 < \kappa < 8$, and $\kappa\ge8$.}
\end{figure}
Moreover, $\kappa $ is related to the fractal dimension of a curve by the following relation,
\begin{equation}
d = 1+ \frac{\kappa}{8}
\end{equation}
Physically, for $0 <\kappa < 4$, the repulsive force wins and the particle escapes to infinity,
while for $\kappa>4$, the noise dominates and the particle collides with the origin in finite
time\cite{Cardy:2005kh}.

Let's turn to the dual aspect of gravity. What we know is that there is Hawking-Page phase transition which is between thermal radiation phase and
the black hole phase in the AdS space. However, we still don't know what the internal structure of three-dimensional AdS is. Fortunately, we can further understand the structure of three-dimensional AdS from the aspect of SLE. Next, we will take the three-dimensional AdS with $c=\frac{1}{2}$ as an example to explain in detail. According to the previous analysis, $\kappa$ has two values at $c=\frac{1}{2}$, namely, $\kappa=3$ or $\kappa=\frac{16}{3}$. It is worth noting that the particles in the AdS space with $\kappa=3$($0 <\kappa \le 4$) are dominated by repulsion, and the particle escapes to infinity. However, when $\kappa=\frac{16}{3}$($4 < \kappa < 8$), the particle clusters in AdS space collide with the origin in final time(see Fig.1)\cite{Cardy:2005kh}. Note that these conclusions cannot be obtained from Hawking-Page transition. And this conclusion can be regarded as an effective supplement to HP phase transition.

In fact, the bulk dual gravity with the central charge $c=1$ is still unknown\cite{Castro:2011zq}. So, it is natural to ask that can we study this bulk dual by SLE. Actually, by simple calculation, we find that $\kappa$ is equal to 4 when $c=1$. Then, from the result of SLE, one can find that $\kappa = 4$ corresponds to the path of the harmonic explorer and contour lines of the Gaussian free field. Physically, the Gaussian free field is also called the bosonic massless free field.

\begin{table}
\begin{center}
\caption{central charge c, CFT, bulk dual, $\kappa$ of SLE are showed. Each value of $c < 1$ corresponds to two values of $\kappa$, one value $\kappa$ between 0 and 4, and the other value $\frac {16}{\kappa}$ greater than 4}

{\begin{tabular}{@{}cccccccccc} \hline
 & \multicolumn{2}{c}{c}&\multicolumn{2}{c}{CFT}&\multicolumn{2}{c}{bulk~dual}&\multicolumn{2}{c}{$\kappa$} \\ \hline
 & \multicolumn{2}{c}{$\frac {1}{2}$}&\multicolumn{2}{c}{Ising ~model}&\multicolumn{2}{c}{pure~gravity}&\multicolumn{2}{c}{$3,\frac{16}{3}$}\\

&\multicolumn{2}{c}{$\frac{7}{10}$}&\multicolumn{2}{c}{tricritical~Ising~model}&\multicolumn{2}{c}{pure gravity}&\multicolumn{2}{c}{$5,\frac{16}{5}$}     \\ 
&\multicolumn{2}{c}{$\frac {4}{5}$}&\multicolumn{2}{c}{Potts~model}&\multicolumn{2}{c}{$SL_{3}$ gravity}&\multicolumn{2}{c}{$\frac{10}{3},\frac{24}{5}$} \\
&\multicolumn{2}{c}{$\frac {6}{7}$}&\multicolumn{2}{c}{tricritical~Potts~model}&\multicolumn{2}{c}{$E_{6}$ gravity}&\multicolumn{2}{c}{$\frac{24}{7},\frac{14}{3}$} \\
&\multicolumn{2}{c}{$2 \frac {N-1}{N+2}$}&\multicolumn{2}{c}{parafermions~model}&\multicolumn{2}{c}{$SL_{N}$ gravity} \\\hline

\end{tabular}
\label{mass_b2}}
\end{center}
\end{table}

\section{Phase Transition in Ising model and three dimensional gravity }\label{sec:03}
Maloney and Witten~\cite{Maloney:2007ud} shown that the Hawking-Page
transition~\cite{Hawking:1982dh} in $AdS_{3}$ space is Lee-Yang type, while the original Lee-Yang phase transition~\cite{Lee:1952ig,Yang:1952be} is only for two dimensional Ising model. The Hawking-Page transition can be seen from Lee-Yang condensation of zeros in the partition function when
$k\rightarrow\infty$. Actually, the partition function $Z(\tau)$ of three dimensional gravity is a modular function which computes at fixed temperature Im$\tau$ and angular potential Re$\tau$. In the limit of infinite volume, these zeroes condense in the phase boundaries. When more and more zeros are gathered, it can undergo phase transition. The analog of the infinite volume limit for the partition function $Z(\tau)$ is $k\rightarrow\infty$. This is because of $k=\ell / 16 G$ implies that $k$ directly proportional to the $AdS_{3}$ radius. In this case, the partition function $Z(\tau)$ is not analytic which corresponding to the occurrence of phase transition. Besides, we know that partition function of pure quantum gravity at $c=\frac {1}{2}$ is equal to that of the Ising mode. So it naturally to study the phase transition of the systems.

The three-dimensional Hawking-Page transition represents  the transition from BTZ black hole with thermal radiation to thermal $AdS_{3}$. And the critical temperature read
\begin{equation}
T_{HP}= \frac {1}{2\pi\ell}.
\end{equation}
On the other hand, the central charge in AdS space can be written as\cite{Castro:2011zq,Santachiara:2007oik}
\begin{equation}
c= \frac {3 \ell}{2 G}
\end{equation}
Note that the three dimensional quantum gravity at $c= \frac{1}{2}$ is dual to the Ising model. Then combining Eq.(12) and Eq.(13) with $c=\frac {1}{2}$, we find
\begin{equation}
T_{HP}= \frac {3}{2\pi G}.
\end{equation}
This means that the Hawking-Page
transition in the Ising model with $c=\frac {1}{2}$ is only determined by three dimensional Newton's gravitational constant $G$.

\section{Summary}\label{sec:04}
In summary, we introduce the three dimensional gravity and the dual Ising model. We argue that the three dimensional gravity can be described by the Schramm-Loewner evolution. SLE is stochastic processes which exhibits conformal invariance.  We also obtain different $\kappa$ for some special model such as pure gravity, $E_{6}$ gravtiy, $SL_{3}$ gravity and $SL_{N}$ gravity\cite{Castro:2011zq}. This shows that the gravity has two different phases.  Since $AdS_{3}$ gravity at $c= \frac{1}{2}$ is dual to the Ising model,
we argue that the Hawking-Page
transition in the Ising model is only determined by three dimensional Newton gravity constant $G$.

For some special central charge c, we show that the three gravity is dual to SLE
at certain $\kappa$, which is the weight of the Brownian motion. Then it is natural to study whether SLE is dual to the three dimensional gravity for a general case when $0<c<1$. And it is worth researching in the future.



\section*{Acknowledgment}
We thank for discussing with Qi An. This work is partly supported by the Research Foundation of Education
Bureau of Hunan Province, China under Grant No. 22B0788 and the Natural Science Foundation of Hunan Province,
China under Grant No. 2021JJ40020.

\section*{Competing Interests Statement}
Competing interests: The authors declare there are no competing interests
\section*{The Data Availability Statement}
Data generated or analyzed during this study are available from the corresponding author upon reasonable request.

\end{document}